\documentclass[12pt,english]{article}
\usepackage{babel}
\usepackage{amsfonts}
\title{Spontaneous generation of the Nambu --Jona-Lazinio 
\\ 
interaction in quantum chromodynamics \\
with two light quarks}
\author{B.A. Arbuzov \\
{\it Skobeltzyn Institute of Nuclear Physics of 
MSU}\\ {\it 119992 Moscow, RF}\\
E-mail: arbuzov@theory.sinp.msu.ru}
\date{}
\newcommand{\be}{\begin{equation}}
\newcommand{\ee}{\end{equation}}
\newcommand{\beq}{\begin{eqnarray}}
\newcommand{\eeq}{\end{eqnarray}}       
\newcommand{\nn}{\nonumber}
\newcommand{\bi}{\bibitem}
\begin{document}
\maketitle
\begin{quote}
 
In QCD with two light quarks with application of 
Bogolubov quasi-averages approach a possibility of 
spontaneous generation of an effective interaction, 
leading to the Nambu -- Jona-Lazinio model, is studied.
Compensation equations for form-factor of 
the interaction is shown to have the non-trivial solution 
leading to theory with two parameters: average low-energy 
value of $\alpha_s$ and dimensional parameter $f_\pi$.
 All other 
parameters: the current and the constituent quark 
masses, the quark condensate, mass of 
$\pi$ meson, mass of $\sigma$ meson and its width  
are expressed in terms of the two 
initial parameters in satisfactory correspondence to 
experimental phenomenology. The results being obtained 
allow to state an applicability of the approach in 
the low-energy hadron physics and promising possibilities 
of its applications to other problems.
\end{quote}

\section{Introduction}

In work~\cite{Arb04} based on quasi-averages approach by 
N.N. Bogolubov~\cite{Bog2, Bog} a method is proposed 
aimed 
to obtain an effective interaction in a  
renormalizable quantum field theory. In particular such 
an effective interaction is necessary for construction of 
the well-known Nambu -- Jona-Lazinio model~\cite{Nambu},  
which describes the low-energy physics of strong 
interactions. In view of this it is of interest to 
apply the developed method~\cite{Arb04} to a study of 
possibility of generation of effective four-fermion 
interaction, which is intrinsic to the Nambu -- Jona-
Lazinio model. Following the results of work~\cite{Arb04} 
one is to expect the essential contraction of number 
of initial parameters of the model.

An application of the method~\cite{Arb04} to a rather  
extensively studied low-energy region of hadron physics 
may show to what extent the use of the method is justified. 
In the present work the first non-perturbative 
approximation of the method will be developed in 
application to the problem of a spontaneous generation 
of Nambu -- Jona-Lazinio Lagrangian in the genuine 
strong interaction theory, namely in QCD with doublet 
of light quarks.

\section{Compensation equation for effective 
form-factor}

Now we start with QCD Lagrangian with two light quarks 
($u$ and $d$) with number of colours $N = 3$
\be
L\,=\,\sum_{k=1}^2\biggl(\frac{\imath}{2}
\Bigl(\bar\psi_k\gamma_
\mu\partial_\mu\psi_k\,-\partial_\mu\bar\psi_k
\gamma_\mu\psi
_k\,\biggr)-\,m_0\bar\psi_k\psi_k\,+
\,g\bar\psi_k\gamma_\mu t^a A_\mu^a\psi_k
\biggr)\,-
\,\frac{1}{4}\,\biggl( F_{\mu\nu}^aF_{\mu\nu}^a\biggr);
\label{initial}
\ee
where we use the standard QCD notations.

In accordance to the Bogolubov approach, application of 
which to such problems being described in details in work
~\cite{Arb04}, we look for a non-trivial solution of a 
compensation equation, which is formulated on the basis 
of the Bogolubov procedure "add -- subtract". Namely 
let us write down the initial expression~(\ref{initial}) 
in the following form
\beq
& &L\,=\,L_0\,+\,L_{int}\,;\nn\\
& &L_0\,=\,\frac{\imath}{2}
\Bigl(\bar\psi\gamma_
\mu\partial_\mu\psi-\partial_\mu\bar\psi\gamma_\mu\psi
\biggr)\,-\,\frac{1}{4}\,F_{0\,\mu\nu}^aF_{0\,\mu\nu}^a\,-
\,m_0\bar\psi\,\psi\,+ 
\,\frac{G_1}{2}\cdot\Bigl(
\bar\psi\tau^b\gamma_5\psi\,\bar\psi \tau^b\gamma_5
\psi\,-\nn\\
& &-\bar\psi\,\psi\,\bar\psi\,\psi\biggr)\,+\,\frac{G_2}
{2}\cdot\Bigl(\bar\psi\tau^b\gamma_\mu\psi\,\bar\psi
\tau^b\gamma_\mu\psi +
\bar\psi \tau^b\gamma_5 \gamma_\mu \psi \bar\psi \tau^b
\gamma_5 \gamma_\mu \psi\biggr)\,;\label{addsub}\\
& &L_{int}\,=\,g_s\,\bar\psi\gamma_\mu 
t^a A_\mu^a\psi\,-\,\frac{1}{4}\,\biggl( F_{\mu\nu}^aF_
{\mu\nu}^a - 
F_{0\,\mu\nu}^aF_{0\,\mu\nu}^a\biggr)\,-\,\frac{G_1}{2}
\cdot\Bigl(\bar\psi \tau^b\gamma_5 \psi\,\bar\psi \tau^b
\gamma_5\psi -\bar\psi\,\psi\,\bar\psi\,\psi\biggr)\,-
\nn\\
& &-\,\frac{G_2}{2}\cdot\Bigl(\bar\psi \tau^b\gamma_\mu 
\psi\,\bar\psi \tau^b\gamma_\mu \psi +
\bar\psi \tau^b\gamma_5 \gamma_\mu \psi \bar\psi \tau^b
\gamma_5 \gamma_\mu \psi\biggr)\,.\label{intas}
\eeq
Here $\psi$ is the isotopic doublet, colour summation is 
performed inside of each fermion 
bilinear combination, $F_{0\,\mu\nu} = 
\partial_\mu A_\nu - \partial_\nu A_\mu$, and notation  
 $G_1\cdot \bar\psi \psi \bar\psi \psi$ corresponds to 
non-local vertex in the momentum space
\be
\imath\,(2\pi)^4\,G_1\,F_1(p1,p2,p3,p4)\,
\delta(p1+p2+p3+p4)\,;
\label{vertex}
\ee
where $F_1(p1,p2,p3,p4)$ is a form-factor and 
$p1,\;p2,\;p3,\;p4$ are incoming momenta. 
In the same way we define vertices, containing 
Dirac and isotopic matrices. We comment the composition 
of the vector sector, being proportional to $G_2$, 
in what follows.

Let us consider  expression~
(\ref{addsub}) as the new {\bf free} Lagrangian $L_0$, 
whereas expression~(\ref{intas}) as the new 
{\bf interaction} Lagrangian $L_{int}$. Then 
compensation conditions (see again~\cite{Arb04}) will 
consist in demand of full connected four-fermion vertices, 
following from Lagrangian $L_0$, to be zero. This demand 
gives a set of non-linear equations for form-factors 
$F_i$.

These equation according to terminology of works
~\cite{Bog2, Bog} are called {\it compensation equations}. 
In a study of these equations it is always evident the 
existence of a perturbative trivial solution (in our case 
$G_i = 0$), but, in general, a non-perturbative 
non-trivial solution may also exist. Just the quest of 
a non-trivial solution inspires the main interest in such 
problems. One can not succeed in finding an exact 
non-trivial solution in a realistic theory, therefore 
the goal of a study is a quest of an adequate 
approach, the first non-perturbative approximation of 
which describes the main features of the problem. 
Improvement of a precision of results is to be achieved 
by corrections to the initial first approximation.

Thus our task is to formulate the first approximation. 
Here the experience acquired in the course of performing 
of work~\cite{Arb04} could be helpful. Now in view of 
obtaining the first approximation we assume:\\
1) In compensation equation we restrict ourselves by 
terms with loop numbers 0, 1, 2.\\
2) In expressions thus obtained we perform a procedure 
of linearizing, which leads to linear integral 
equations. It means that in loop terms only one vertex 
contains the form-factor, being defined above, while 
other vertices are considered to be point-like. In
diagram form equation for form-factor $F_1$ is presented 
in Fig 1.\\
3) While evaluating diagrams with point-like vertices 
diverging integrals appear. Bearing in mind that as a 
result of the study we obtain form-factors decreasing at 
momenta infinity, we introduce a cut-off $\Lambda$ in 
the diverging integrals. It will be shown that results do 
not depend on the value of this cut-off.\\
4) We can obtain analytic expressions for massless 
quarks only. We take masses into account by introducing 
the cut-off in the lower limit of integration by 
momentum squared $q^2$ at a value, which equals to 
a value of the corresponding propagator denominator 
at $q = 0$. Namely with the denominator $(q^2+m^2)$ 
the cut-off parameter equals $m^2$. In doing this we 
keep at nominators only the leading terms in $m$ 
expansions because taking into account of the next 
terms evidently means supererogation of accuracy. \\
5) We shall take into account at most the first two 
terms of the $1/N$ expansion.

Let us formulate compensation equations in this 
approximation. 
For {\it free} Lagrangian $L_0$ full connected 
four-fermion vertices are to vanish. One can succeed in 
obtaining analytic solutions for the following set 
of momentum variables (see Fig. 1): left-hand legs 
have momenta  $p$ and $-p$, and right-hand legs 
have zero momenta. In particular this kinematics 
suits for description of zero-mass bound states.  
Under some assumptions solutions obtained under these 
conditions may be generalized to momentum set 
$(p,-p,q,-q)$. The construction of expressions with 
an arbitrary set of momenta is the problem for the 
subsequent approximations. 

Now following the rules being stated above we 
obtain the following equation for form-factor $F_1(p)$ in 
scalar colour singlet channel
\beq
& &G_1 F_1(p^2) = 
\frac{G_1^2 N \Lambda^2}{2 \pi^2}\biggl(1 + \frac{1}{4 N}
-\frac{G_1 N }{2 \pi^4}\Bigl(1 + \frac{1}{2 N}\Bigr)
\int \frac{F_1(q^2)\,dq}{q^2}\,\biggr)+
\nn\\
& &+\,\frac{3 G_1\, G_2}{8 \pi^2}\biggl
(2 \,\Lambda^2 + p^2 \log\frac{p^2}
{\Lambda^2}-\frac{3}{2}\, p^2 - \frac{\mu^2}{2\,p^2} 
\biggr)\, -\,\frac{(G_1^2+6 G_1 G_2) N}{32\, \pi^6}
\times\nn\\
& &
\times \int\Bigl(
2 \Lambda^2+(p-q)^2 \log\frac{(p-q)^2}{\Lambda^2}-\frac{3}
{2} (p-q)^2 - \frac{\mu^2}{2 (p-q)^2}
\biggr) \frac{ G_1 F_1(q^2) dq}{q^2}\,
;\label{eq1}
\eeq 
Here integration is performed in the four-dimensional 
Euclidean momentum space, $\mu = m_0^2$. One-loop terms 
contains terms 
proportional to $N$ and $1$ while two-loop terms 
correspond to $N^2$ and $N$. The leading terms are the 
same for scalar and pseudo-scalar cases. 
We begins the study with the scalar channel, because 
it defines chiral symmetry breaking effect. 
Equation~(\ref{eq1}) evidently has trivial solution 
$G_1 = 0$. Bearing in mind our goal to look for 
non-trivial solutions we divide the equation by $G_1$. 
As a result of four-dimensional angle integration 
(see Appendix) we have
\beq
& &F_1(x) = A + \frac{3\,G_2}{8 \pi^2}\biggl  
(2\Lambda^2 + x \log\frac{x}
{\Lambda^2}-\frac{3}{2}\, x -\frac{\mu^2}{2 x}\biggr) - 
\frac{(G_1^2+6 G_1 G_2) N}{32\, \pi^4}\times\nn\\
& &\times \Biggl(\frac{1}{6\,x}
\int_\mu^x (y^2-3 \mu^2) F_1(y)\,dy\,+\,\frac{3}{2}
\int_\mu^x y F_1(y)\,dy + 
\log\,x \int_\mu^x y F_1(y)\,dy +\nn\\
& &+ x\,\log\,x \int_\mu^x 
F_1(y)\,dy + 
\int_x^\infty y \log\,y\, F_1(y)\,dy +\,x \int_x^\infty  
\Bigl(\log\,y\,+\frac{3}{2}\biggr) F_1(y)\,dy\,+\nn\\
& &+\,\frac{x^2-3 \mu^2}{6}
\int_x^\infty\,\frac{F_1(y)}{y}\,dy + \Bigl(2\Lambda^2-
\frac{3}{2}\, x\Bigr)\int_\mu^\infty F_1(y)\,dy - 
\frac{3}{2}\,\int_\mu^\infty y F_1(y)\,dy\,-\nn\\
& &-\,\log\,
\Lambda^2 \Bigl(\int_\mu^\infty y F_1(y)\,dy\,+\,
x \int_\mu^\infty F_1(y)\,dy\Bigr)\,\Biggr)\,;\label{eq1xy}
\\
& &A\, = \,
\frac{G_1^2 N \Lambda^2}{2 \pi^2}\biggl(1 + \frac{1}{4 N}
-\frac{G_1 N }{2 \pi^2}\Bigl(1 + \frac{1}{2 N}\Bigr)
\int_\mu^\infty F_1(y)\,dy\,\biggr)\,;\nn\\
& &x\,=\,p^2\,;\qquad y\,=\,q^2\,.\nn  
\eeq
In view of taking into account of quark mass the 
lower limit of the momentum integration $\mu = m_0^2$ 
is introduced. Equation~(\ref{eq1xy}) by a sequential 
six-fold differentiation reduces to the following 
differential equation
\beq
& &\frac{d^2}{dx^2}\Biggl(x\,\frac{d^2}{dx^2}\biggl(x\,
\frac{d^2}{dx^2}\Bigl(x\,F_1(x)\biggr)+\frac{\beta\,
\mu^2}{4}\,F_1(x)\biggr)\Biggr)\,=\,
\beta\,\frac{F_1(x)}{x}\,.\label{diff1}\\
& &\beta\,=\,\frac{(G_1^2+6\,G_1 G_2)\,N}{16\,\pi^4}\,;
\nn
\eeq
with boundary conditions to be formulated below.

Equation~(\ref{diff1}) reduces to Meijer 
equation~\cite{be},~\cite{Mar3}. Namely with the simple 
substitution we have
\beq
& &\Biggl(\biggl(z\,\frac{d}{dz}-b\biggr)\biggl(z\,\frac
{d}{dz}-a\biggr)\,z\,\frac{d}{dz}\biggl(z\,\frac{d}{dz}-
\frac
{1}{2}\biggr)\biggl(z\,\frac{d}{dz}-\frac{1}{2}\biggr)
\biggl
(z\,\frac{d}{dz}-1\biggr) - z\Biggr)
 F_1(z) =\,0\,;\label{Meijer}\\
& &z\,=\,\frac{\beta\,x^2}{2^6}\,;\quad a\,=\,-\,\frac{1-
\sqrt{1- 64 u}}{4}\,;\quad 
b\,=\,-\,\frac{1+\sqrt{1- 64 u}}{4}\,;\quad 
u\,=\,\frac{\beta\,\mu^2}{64}\,.\nn
\eeq
Boundary conditions for equation~(\ref{Meijer}) are 
formulated in the same way as in works~\cite{Arb04}, 
\cite{AF}. At first we have to choose solutions 
decreasing at infinity, that is combination of the 
following three solutions 
\beq
& &F_1(x)\,=\,C_1\,G_{06}^{40}\Bigl(z\,
|1,\,\frac{1}{2},\,\frac{1}{2},\,0,\,a,\,
b\Bigr)\,+\,C_2\,G_{06}^{40}\Bigl(z\,|1,\,\frac{1}{2},\,b,
\,a,\,\frac{1}{2},\,0,\,\Bigr)\,+\nn\\
& &+\,C_3\,G_{06}^{40}\Bigl(z\,
|1,\,0,\,
b,\,a,\,\frac{1}{2},\,\frac{1}{2}\Bigr)\,;\qquad z\,=\,
\frac{\beta\,x^2}{2^6}\,.\label{Ci}
\eeq
where
$$
G^{m n}_{p q}\biggl( z\,|^{a_1,\,...\,,\,a_p}_
{b_1,\,...\,,\,b_q}\biggr)\,;
$$
is the Meijer function~\cite{be},~\cite{Mar3} with 
sets of upper indices ${a_i}$ and of lower ones ${b_j}$. 
In case only one line of parameters is written this 
means the presence of lower indices only, $n$ and $p$ in 
the case being equal to zero.
 
Constants $C_i$ are defined by conditions
\beq
& & \frac{3\,G_2}{8 \pi^2}-\frac{\beta}{2}
\int_\mu^\infty F_1(y)\,dy\,=\,0\,;\nn\\
& &\int_\mu^\infty y\,F_1(y)\,dy\,=\,0\,;\label{bound}\\
& &\int_\mu^\infty y^2\,F_1(y)\,dy\,=\,0\,.\nn
\eeq

These conditions and condition $A = 0$ as well provide 
cancellation of all terms in Eq.~(\ref{eq1xy}) being 
proportional to $\Lambda^2$ and $\log\,\Lambda^2$. Thus 
the result does not depend on a value of parameter 
$\Lambda$. 
By solving linear set~(\ref{bound}), in which 
solution~(\ref{Ci}) is substituted, we obtain the unique 
solution. Value of parameter $u_0$, which is connected 
with current quark mass, and ratio of two constants $G_i$ 
we obtain from conditions $F_1(\mu)=1$ and
\beq
& &A\,=\,\frac{G_1 N \Lambda^2}{2 \pi^2}\biggl(1+ \frac{1}
{4 N} -\,\frac{G_1 N }{2 \pi^2}\Bigl(1 + \frac{1}{2 N}
\Bigr)\int_\mu^\infty F_1(y)\,dy\,\biggr)\,=\nn\\
& &=\,\Bigl(1+ \frac{1}
{4 N}\Bigr)\frac{G_1 N \Lambda^2}{2 \pi^2}\biggl(1- \frac{6 G_2 (4 N+2)}
{(G_1 + 6 G_2)(4 N+1)}\biggr)\,=\,0\,;\label{A0}
\eeq
this gives for $N=3$ with the account of the first of 
conditions~(\ref{bound})
\be
u_0\,=\,1.6\cdot 10^{-8}\,;\quad 
G_1\,=\,\frac{6}{13}\,G_2\,.\label{result1}
\ee

The form-factor now reads as ~(\ref{Ci}) with 
\be
C_1\,=\,0.28322\,;\quad C_2\,=\,-\,3.655\cdot10^{-8}\,;
\quad C_3\,=\,-\,7.794\cdot10^{-8}\,;
\ee
$F_1(u_0) = 1$ and $F_1(z)$ decreases with $z$ increasing.
It is important, that the solution exists only for 
positive $G_2$ and due to~(\ref{result1}) for positive 
$G_1$ as well. 

Let us comment an essential point, connected with very 
small value of parameter $u_0$. Note, that for $u = 0$ 
solution satisfying all conditions excluding~(\ref{A0}) is given by the following expression
$$
\frac{1}{2\,\sqrt{\pi}}\;G_{06}^{40}\Bigl(z\,
|1,\,\frac{1}{2},\,\frac{1}{2},\,0,\,0,\,
-\,\frac{1}{2}\Bigr)\,;
$$
with ratio
\be
r(u)_{u = 0}\,=\,G_1/G_2\,=\,-\,3\,+\sqrt{12}\,=\,
0.464102\,.\label{x0}
\ee
At the same time value~(\ref{result1}), which follows from 
~(\ref{A0}) is equal to  $0.461538$. For $u \to 0$ 
this ratio $x(u)$
tends to value~(\ref{x0}) from below. Because of these 
two numbers being quite close the intersection of 
curve $x(u)$ with~(\ref{A0}) occurs at so small $u$. 
In case we use in expression~(\ref{A0}) the most 
leading terms in $1/N$, the ratio is $1/2$ and there is 
no intersection of $x(u)$ with this value at all, i.e. 
the solution does not exist. So a value $u_0$ is 
sharply dependent on approximations and quite small 
change of coefficients in expression~(\ref{A0}) 
influences it strongly. We take into account these 
considerations while commenting values of current quark 
mass in what follows.

Solution $F_1(p,-p,0,0)$ can be extended for 
more general kinematics. Namely, let us consider 
form-factor $F_1(p,-p,q,-q)$. Assuming factorization 
property and bearing in mind 
the previous note on small contribution of $u_0$ the  
form-factor reads as follows
\beq
& &F_1(p,\,-p,\,q,\,-q) = \frac{1}{4\,\pi}\,G_{06}^{40}
\Bigl(z\,
|1,\,\frac{1}{2},\,\frac{1}{2},\,0,\,0,\,
-\,\frac{1}{2}\Bigr)\,G_{06}^{40}\Bigl(z'\,
|1,\,\frac{1}{2},\,\frac{1}{2},\,0,\,0,\,
-\,\frac{1}{2}\Bigr)\,;\nn\\
& &z\,=\,\frac{\beta\,x^2}{64}\,;\quad z'\,=\,\frac{\beta
\,y^2}{64}\,; \quad x\,=\,p^2\,;\quad y\,=\,q^2\,.
\label{pq}
\eeq  
A study of form-factor $F_2(p^2)$, which enters into 
four-fermion vector and pseudo-vector terms 
in~(\ref{addsub}), leads to result, that conditions of an 
existence  of the corresponding solution does not provide 
additional restrictions for parameters being introduced 
above. Vector form-factors will be considered in details 
elsewhere. The explicit form of $F_2(p^2)$ do not 
influence results of the present work.

\section{Wave function for scalar and pseudo-scalar 
excitations}
Now with the non-trivial solution of the compensation 
equation we arrive at an effective theory in which
there are already no  undesirable four-fermion terms in 
{\bf free} Lagrangian~(\ref{addsub}) while they are 
evidently present in {\bf interaction} Lagrangian
~(\ref{intas}). Indeed four-fermion terms in these two 
parts of the full Lagrangian differ in sign and 
the existence of the non-trivial solution of 
compensation equation for Lagrangian~(\ref{addsub}) 
means non-existence of the would be analogous equation, 
formulated for signs of four-fermion terms in 
{\bf interaction} Lagrangian~(\ref{intas}). 
In other words the fact, that sum of a series 
$\sum G^n a_n = 0$, by no means leads to a conclusion, 
that sum of the same series with $G \to -\,G$ vanishes as 
well.

So provided the non-trivial solution  
is realized the compensated 
terms go out from 
Lagrangian~(\ref{addsub}) and we obtain the following 
Lagrangian 
\beq
& &L\,=\,\frac{\imath}{2}
\Bigl(\bar\psi\gamma_
\mu\partial_\mu\psi-\partial_\mu\bar\psi\gamma_\mu\psi
\biggr)\,-\,\frac{1}{4}\,F_{0\,\mu\nu}^aF_{0\,\mu\nu}^a\,-
\,m_0\bar\psi\,\psi\,+ \nn\\
& &+\,g_s\,\bar\psi\gamma_\mu 
t^a A_\mu^a\psi\,-\,\frac{1}{4}\,\biggl( F_{\mu\nu}^aF_
{\mu\nu}^a - 
F_{0\,\mu\nu}^aF_{0\,\mu\nu}^a\biggr)\,-\,\frac{G_1}{2}
\cdot\Bigl(\bar\psi \tau^b\gamma_5 \psi\,\bar\psi \tau^b
\gamma_5\psi -\bar\psi\,\psi\,\bar\psi\,\psi\biggr)\,-
\nn\\
& &-\,\frac{G_2}{2}\cdot\Bigl(\bar\psi \tau^b\gamma_\mu 
\psi\,\bar\psi \tau^b\gamma_\mu \psi +
\bar\psi \tau^b\gamma_5 \gamma_\mu \psi \bar\psi \tau^b
\gamma_5 \gamma_\mu \psi\biggr)\,.\label{intas1}
\eeq
Thus, bound state problems in the present approach are 
formulated starting from Lagrangian (\ref
{intas1}).

In case of realization of the nontrivial solution, let 
us consider zero-mass scalar and pseudo-scalar 
excitations . 
Let us write down Bethe-Salpeter 
equation for scalar zero-mass channel in the same 
approximation as was used in equation~(\ref{eq1xy})
\beq
& &\Psi(x) = \frac{G_1\,N}{2 \pi^2}\int_\mu^
{\infty} 
\Psi(y)\,dy +  \frac{(G_1^2+6 G_1 G_2) N}
{32\, \pi^4} \Biggl(\frac{1}{6\,x}\int_\mu^x 
(y^2 - 3 \mu^2) \Psi(y)\,
dy\,+\nn\\
& &+ \frac{3}{2}\int_\mu^x y \Psi(y)\,dy + 
\log\,x \int_\mu^x y \Psi(y)\,dy + x\,\log\,x \int_\mu^x 
\Psi(y)\,dy + 
\int_x^{\infty} y \log\,y\, \Psi(y)\,dy +\nn\\
& &+\,x \int_x^{\infty}  
\Bigl(\log\,y\,+\frac{3}{2}\biggr) \Psi(y)\,dy\,+\,
\frac{x^2 - 3 \mu^2}{6}
\int_x^{\infty}\,\frac{\Psi(y)}{y}\,dy + 
\Bigl(2\bar\Lambda^2-
\frac{3}{2}\, x\Bigr)\int_\mu^{\infty} \Psi(y)\,dy - \nn\\
& &\frac{3}{2}\,\int_\mu^{\infty} y \Psi(y)\,dy\,-\,\log\,
\Lambda^2 \Bigl(\int_\mu^{\infty} y \Psi(y)\,dy\,+\,
x \int_\mu^{\infty} \Psi(y)\,dy\Bigr)\,\Biggr)\,;
\label{eq1psi}
\nn
\eeq
The corresponding differential equation for $\Psi(x)$ is 
almost the same, as the previous one~(\ref{diff1}) with 
one essential difference. Namely the sign afore $\beta$ 
is opposite.
\beq
& &\Biggl(\biggl(z \frac{d}{dz}-b\biggr)\biggl(z \frac
{d}{dz}-a\biggr)\,z \frac{d}{dz}\,\biggl(z \frac{d}{dz}-
\frac{1}{2}\biggr)\biggl(z \frac{d}{dz}-\frac{1}{2}\biggr)
\biggl
(z \frac{d}{dz}-1\biggr) + \frac{\beta\,z}{2^6}\Biggr)
\,F(z)\,=\,0\,;\label{Meijerpsi}\\
& &z\,=\,x^2\,;\quad a\,=\,\frac{-1+\sqrt{1-64 u}}{4}\,;
\quad b\,=\,\frac{-1-\sqrt{1-64 u}}{4}\,;\quad 
u\,=\frac{\beta m^4}{64}\,.
\eeq
Boundary conditions read
\be
\int_\mu^{\infty} \Psi(y)\,dy\,=\,0\,;\quad 
\int_\mu^{\infty} y\,\Psi(y)\,dy\,=\,0\,;\quad 
\int_\mu^{\infty} y^2\,\Psi(y)\,dy\,=\,0\,.\label{boundpsi}
\ee
Now we have four independent solutions decreasing at 
infinity, which we use for general solution
\beq
& &\Psi(x)\,=\,C^*\,G_{06}^{30}\Bigl(\frac{
\beta\,
x^2}{2^6}|1,\,\frac{1}{2},\,0,\,\frac{1}{2},\,a,\,
b\Bigr)\,+\,C_1^*\,G_{06}^{30}\Bigl(\frac{
\beta\,
x^2}{2^6}|1,\,\frac{1}{2},\,\frac{1}{2},\,0,\,a,\,
b\Bigr)\,+\nn\\
& &+\,C_2^*\,G_{06}^{30}\Bigl(\frac{
\beta\,
x^2}{2^6}|1,\,a,\,
b,\,\frac{1}{2},\,\frac{1}{2},\,0\Bigr)+\,C_3^*\,
G_{06}^{30}\Bigl(\frac{\beta\,
x^2}{2^6}|\frac{1}{2},\,a,\,b,\,1,\,\frac{1}{2},\,0\Bigr)
\,.\label{cpsi}
\eeq
Boundary conditions~(\ref{boundpsi}) allows to express 
$C_i^*,\,i =1,2,3$ in terms of $C^*$ and condition of 
wave function to be unity on the mass shell $\Psi(m^2)\,
=\,1$ fixes $C^*$ as well. 
The normalization condition of the Bethe-Salpeter wave 
function fixes constant of interaction of the bound state 
with quark-anti-quark pair 
\be
g\,\Bigl(\phi\,\bar\psi\,\psi+\imath\,\pi_a\,\bar\psi\,
\gamma_5\,\tau_a\psi\Bigr)\,.\label{sigpsi}
\ee
Namely this normalization condition, which 
corresponds to a correct coefficient afore 
a kinetic part of scalar and pseudo-scalar 
particles, gives
\be
\frac{g^2 \,N}{4\,\pi^2}\,I_1\,=\,1\,; \quad I_1\,=\,
\int_\mu^\infty 
\frac{\Psi(x)^2\,dx}{x}\,;\label{g}
\ee
The loop integral of the fourth order in 
interaction~(\ref{sigpsi}) gives four-fold 
interaction  
\be
\lambda\,\frac{(\phi^2\,+\,\pi^a \pi^a)^2}{4!}\,;
\ee
where
\be
-\,\lambda\,=\,\frac{g^4\,3\,N}{\pi^2}\,I_2\,; \quad 
I_2\,=\,\int_\mu^\infty \frac{\Psi(x)^4\,dx}{x}\,
.\label{lambda}
\ee

The result, that equation~(\ref{eq1psi}) has unique 
solution, which satisfies all boundary conditions, is 
the confirmation of the evident fact, that in the same 
approximation, in which the compensation equation has a 
non-trivial solution, fields $\phi$ and $\pi_a$ are to 
have zero masses in accordance with Bogolubov-Goldstone 
theorem~\cite{Bog2, Bog, Gold}.
However an account of chromodynamic interaction leads to 
additional contribution to the masses. Let us calculate 
a mass correction term due to QCD interaction.
For the purpose let us take into account terms of the 
first order in $P^2$, where $P$ is the momentum of 
a scalar (and pseudo-scalar) meson and one-loop 
QCD term. We have
\beq
& &\Psi(p^2)\,=\,\frac{G_1\,N}{2\,\pi^4}\,\int
\frac{ \Psi(q^2)
\,dq}{q^2}\biggl(1 - \frac{3\,P^2}{4\,q^2} + 
\frac{(q P)^2}
{(q^2)^2} \biggr)\, 
+\,\frac{(G_1^2+6 G_1 G_2) N}{32\, \pi^6}\,\times\nn\\
& & \int\biggl(
2 \Lambda^2+(p-q)^2 \log\frac{(p-q)^2}{\Lambda^2}-\frac{ 3}{2} (p-q)^2
\biggr)\biggl(1 - \frac{3\,P^2}{4\,q^2} + \frac{(q P)^2}
{(q^2)^2} \biggr)\frac{ \Psi(q^2)\,dq}{q^2}\,+\nn\\
& &+\,\frac{g_s^2}{4\,\pi^4}\,\int \frac{ \Psi(q^2)\,dq}
{q^2 (q-p)^2}\,.\label{eqpsi1}
\eeq
In the course of QCD term calculation we use 
transverse Landau gauge \footnotemark[1]
\footnotetext[1]{In the approximation used the transverse 
gauge leads to absence of renormalization of both 
vertex and spinor field}. Let us multiply equation~(\ref
{eqpsi1}) by $\Psi(p^2)/p^2$ at $P = 0$ and integrate by
$ p $. Due to equation~(\ref{eq1psi}) be satisfied we  
have
\beq
& &-\,\frac{P^2}{2}\,\int \frac{ \Psi(q^2)^2\,dq}{(q^2)^2}
\,+\,\frac{g_s^2}{4\,\pi^4}\,\int \frac{ \Psi(p^2)\,dp}
{p^2} \int \frac{ \Psi(q^2)\,dq}{q^2 (q-p)^2}\,=\,0\,;
\label{P2}
\eeq
After angle integration we get
\beq
& &\frac{P^2\,\pi^2}{2}\,I_1\,=\,\frac{g_s^2}{4}\,
\int_{m^2}^\infty\,\Psi(x)\,dx\,\biggl(\frac{1}{x}\int_
{m^2}^x\,\Psi(y)\,dy\,+\,\int_x^\infty
\frac{\Psi(y)\,dy}{y}\,\biggr)\,=\nn\\
& &=\,\frac{g_s^2}{\sqrt{\beta}}\,
\int_u^\infty\,\,\frac{\Psi(z)\,dz}{z}\int_{u}
^z\,\frac{\Psi(t)\,dt}{\sqrt{t}}\,=\,\frac{g_s^2\,I_3}
{\sqrt{\beta}};
\quad z\,=\frac{\beta\,x^2}{64}\,;\quad 
t\,=\frac{\beta\,y^2}{64}\,.\label{eigen}
\eeq
The integral entering into~(\ref{eigen}) looks like
\beq
& &\int_u^z \frac{\Psi(t)\,dt}{\sqrt{t}}\,=\,\Biggl(C^*\,
G_{06}^{30}\Bigl(
t\,|_{\frac{1}{2},\,1,\,\frac{3}{2},\,0,\,a',\,
b'}\Bigr)\,+\,C_1^*\,G_{17}^{31}\Bigl(t\,|^ 1_{1,\,1,\,
\frac
{3}{2},\,0,\,\frac{1}{2},\,a',\,b'}\Bigr)\,+\nn\\
& &+\,C_2^*\,G_{06}^{30}\Bigl(t\,|_{\frac{3}{2},\,a',\,
b',\,0,\,\frac{1}{2},\,1}\Bigr)+\,C_3^*\,
G_{06}^{30}\Bigl(t\,|_{1,\,a',\,b',\,0,\,\frac{1}{2},\,
\frac{3}{2}}\Bigr)\Biggr)^z_u\,;\label{cpsiint}\\
& &a'\,=\,\frac{1-\sqrt{1-64 u}}{4},\quad 
b'\,=\,\frac{1+\sqrt{1-64 u}}{4}\,.\nn
\eeq
Relation~(\ref{eigen}) gives us mass of scalar (and 
pseudo-scalar) in terms of average low-energy 
QCD constant $\bar\alpha_s$. 
There are a number of considerations concerning 
possible values of the parameter. For example, 
in approach~\cite{sim} with low-energy freezing 
of the strong constant value $\bar\alpha_s = 
0.414$ is quoted. Smooth matching of perturbative and 
non-perturbative regions in QCD gives value $\bar\alpha_s 
= 0.4354$~\cite{aral}. There is also a sum rule definition 
of this parameter from experimental data $\bar\alpha_s = 
0.47 \pm 0.07$~\cite{dok}. there are also considerations 
on behalf of larger values of $\bar\alpha_s$~\cite{braz, 
faust}. Taking into account these remarks we consider 
range of values of the effective low-energy constant 
$\bar\alpha_s\,=\,0.4\,-\,0.75$. In other words we can 
define the same quantity as value of running strong 
constant at a characteristic momentum, e.g. $600\,$MeV. 
Different variants of low energy behaviour of $\alpha_s$ 
lead to the same interval.
Thus we have
\be
m^2_t\,=\,-\,\frac{8\,\bar\alpha_s\,I_3}{\pi \sqrt{\beta}
\,I_1}\,;\quad I_3\,=\,\int_u^\infty\,\frac{\Psi(z)\,
\Psi_1(z)\,dz}{z}\,;\label{mt}
\ee
where $I_1$ is defined in~(\ref{g}). Due to both 
integrals being positive we evidently have tachyon mass 
and so a scalar condensate appears in the minimum of 
effective potential 
$$
m_t^2\,\frac{\phi^2}{2}\,+\,\lambda\,\frac{\phi^4}{24}\,.
$$
For the present approach it is highly important, that the 
account of the QCD interaction leads to tachyon mass of 
a scalar and pseudo-scalars. The appearance of tachyons and 
thus the appearance of the scalar condensate in the 
minimum of the effective potential results in 
stability of the non-trivial solution and consequently 
one may conclude, that Lagrangian~(\ref{intas1}) with 
the conditions being obtained are valid.

Bearing in mind definitions of quantities entering 
in the effective potential, we obtain for value 
of the scalar condensate $\eta$
\be
\eta^2\,=\,\frac{-\,6\, m_t^2}{\lambda}\,; \quad \eta\,=\,
<\phi>\,.\label{eta}
\ee
Mass of $\phi$ after symmetry breaking is 
$\sqrt{2(-\,m_t^2)}$, 
and $\pi$ mass equals to zero in the present approximation.
The constituent quark mass $m_q$ is defined by relation
\be
m_q\,-\,m_0=\,g\,\eta\,=\,g\,\sqrt{\frac{6\,(-\,m_t^2)}{\lambda}}
\,=\,\frac{m_\phi}{2}\,\sqrt{\frac{I_1}{I_2}}\,.\label
{u1/4}
\ee

Integrals in this as well as in other relations of this 
section including boundary conditions are calculated with 
the lower limits defined not by the current mass $m_0$, 
but by the constituent mass $m$. Now parameter $u$ is 
defined by relation
\be
u\,=\,\frac{\beta\,m^4}{2^6}\,.\label{u}
\ee
Thus after the appearance of the scalar condensate the 
quark propagator takes form
\be
G(p)\,=\,\frac{1}{(\gamma\, p)\,-\,\Sigma(p)}\,;\quad 
\Sigma(p)\,=\,(m - m_0)\,\Psi(p^2)\,+\,m_0\,.
\ee
Relations~(\ref{u1/4}, \ref{eigen}, \ref{cpsiint})  
give for $\bar\alpha_s\,=\,0.46$
\beq
& &u\,=\,0.001\,;\quad
C^*\,=\,-\,0.919\,;\quad
C^*_1\,=\,-\,0.0255\,;\quad
C^*_2\,=\,-\,1.895\,;
\nn\\
& &C^*_3\,=\,0.228\,;\quad I_1\,=\,1.285\,;\quad
I_2\,=\,0.801\,;\quad
I_3\,=\,0.302\,.\quad \label{solutu}
\eeq

We take $\bar\alpha_s$ and current mass $m_0$ as the 
initial parameters and express in their terms all 
other parameters including $\pi$-decay constant 
$f_\pi$. We use for the latter the Goldberger -- Treiman 
relation. In the framework of the present approach we 
obtain the relation by considering the expression for 
transition $\pi^+ \to \mu^+\,\nu_\mu$. We have 
\beq
& &f_\pi\,=\,\frac{g\,N}{4\,\pi^2}\int_{m^2}^\infty 
\Bigl((m-m_0)\,\Psi(y)^2\,+\,m_0\,\Psi(y)\Bigr)
\frac{dy}{y}\,=\nn\\
& &=\,\frac{g\,N}{4\,\pi^2}\Bigl((m-m_0)\,I_1
+\,m_0\,I_7\Bigr)\,;\quad I_7\,=\,\int_u^\infty\frac
{\Psi(z)\,dz}{2\,z}\,.\label{fpi}
\eeq
Provided either $m_0 = 0$ or $I_1 = I_7$ we get 
with account of normalization condition~(\ref{g}) just 
the original Goldberger -- Treiman relation 
$m = g\,f_\pi$. We use full relation~(\ref{fpi}). However 
let us note, that values of the two integrals  are close 
$I_1 \simeq I_7$ and the simple original relation works 
with sufficient accuracy.

At this place it is worth-while to explain that we have a 
solution for any value of $m_0$ and the problem is, 
which value is to be chosen. While $m_0$ is not 
measured directly, the accuracy of its definition 
is not very high. For comparison with experimental 
data we propose to fix value of $f_\pi = 93\,$MeV, which  
is quite certain, while 
other parameters including $m_0$ being subjects for 
calculation.

As a result we obtain e.g. for $\bar \alpha_s\,=\,
0.46\;(u = 0.001)$
\be
g\,=\,3.2\,;\;
G_1\,=\,\frac{1}{(240.5\,\mbox{MeV})^2}\,;\;
m\,=\,298.5\,\mbox{MeV}\,;\;
m_0\,=\,19.8\,\mbox{MeV}\,.\label{gmG}
\ee

Mass of  $\sigma$ is defined by above 
results~(\ref{solutu}, \ref{gmG}) 
\be
m_\sigma\,=\,2\,(m - m_0)\,\sqrt{\frac{I_2}{I_1}}\,
=\,440.1\, \mbox{MeV}\,.\label{msig}
\ee
At this stage mass of $\pi$-meson is zero due to 
appearance of vacuum average of scalar field  $\phi$. 

Coupling constant $g_{\sigma\,\pi\,\pi}$ in interaction
\be
g_{\sigma\,\pi\,\pi}\,\sigma\,\pi^a\,\pi^a\,;
\ee
is calculated with the use of wave function~(\ref{cpsi}) 
with parameters~(\ref{solutu}, 
\ref{gmG}). The triangle diagram gives
\be
g_{\sigma\,\pi\,\pi}\,=\,\frac{g^3\,N}{\pi^2}\,
\int_{m^2}^\infty\frac{\Psi(y)^3\,dy}{y}
((m - m_0)\Psi(y) + m_0)\,=\,8.04\,m\,.
\label{gsigma}
\ee
According to~(\ref{gsigma}) the $\sigma$ width reads
\be
\Gamma_\sigma\,=\,\frac{3\,g_{\sigma\,\pi\,\pi}^2}
{16\,\pi\,m_\sigma^2}\,\sqrt{m^2_\sigma - 4 m_\pi^2}\,=\,
605.8\,
\mbox{MeV}\,;\quad m_\pi = 139\,\mbox{MeV}\,.
\label{width} 
\ee
Values~(\ref{msig}, \ref{width}) do not contradict the 
existing data~\cite{pdg}, which, as a matter of fact, 
are characterized by a wide spread of results.

The results on the wave function of a scalar particle 
and on the form-factor of the effective four-fermion 
interaction allow to estimate the quark condensate 
in the next approximation. Indeed vertex 
$\sigma \bar \psi\,\psi$ form-factor is just 
$g \Psi(q^2)\,$. The non-perturbative part of the quark 
propagator reads $g\,\eta\,\Psi(q^2)\,=\,
(m - m_0)\Psi(q^2)\,$.
According to definition of the quark condensate we have
\be
<\bar q\,q>\,=\,-\,\frac{4\,N\,g\eta}{(2\,\pi)^4}\int_
{m^2}^\infty \,\Psi(q^2)\,dq^2\,.\label{qq}
\ee
Here the integral equals zero due to boundary 
conditions~(\ref{boundpsi}). However one may try to 
calculate subsequent approximations. Let us calculate 
loop corrections: firstly with vertex corresponding to 
four-fermion interaction in Lagrangian~(\ref{intas1}) 
and secondly with gluon exchange in transverse gauge 
(see Fig. 3). In loop diagrams with the form-factor 
$F_1(p)$ we sum up infinite loop chain (the senior 
orders in $1/N$), that gives factor $(1 - B)$ in the 
denominator of the first term in the following expression
\beq
& &<\bar q\,q>\,=\,-\frac{12 \sqrt{3}(m - m_0)}{\pi^2 
\sqrt{14}\,G_1\,(1 - B)}
\,I_4\,I_5\,-\,\frac{3\,\bar \alpha_s (m - m_0)}{4\,\pi^3}
\Biggl(-\,
m^2\,I_7\,+\,\frac{16\,\pi^2}{G_1\,\sqrt{42}}\,I_6\Biggr)
\,;\nn\\
& &I_4\,=\,\int_u^\infty \frac{F_1(z)\,dz}{2 \sqrt{z}}\,;
\quad I_5\,=\,\int_u^\infty \frac{F_1(z)\,\Psi(z)\,dz}
{2 \sqrt{z}}\,;\quad
 I_6\,=\,\int_u^\infty \frac{ \Psi_1(z)\,dz}{2 z}\,;
\label{bqq}\\
& & B\,=\,\frac{16\,\sqrt{3}}{\sqrt{14}}\int_u^\infty
\frac{(F_1(z))^2\,dz}{2\,\sqrt{z}}\,;
\nn
\eeq
where function $ \Psi_1(z)$ is defined by 
relation~(\ref{cpsiint}).
We have e.g. again for $u = 0.001,\,\bar 
\alpha_s = 0.46$ 
\be
<\bar q\,q>\,=\,-\,(167\,\mbox{MeV})^3\,;
\ee
that is sufficiently lower than the traditional value 
$(240$ MeV$)^3$. Nevertheless we draw attention to the 
correct sign of the condensate. It is connected with the 
sign afore $G_1$ in the interaction Lagrangian
~(\ref{intas1}),which is defined by conditions of the 
existence of the non-trivial solution of the 
compensation equation.

\section{Discussion of results}

The low-energy hadron parameters being calculated for 
three values of $\bar \alpha_s$ are presented in 
Table 1. The calculations needs numerical integrations 
of expressions containing Meijer functions with infinite 
upper limits. Therefore the presented data may contain 
uncertainties (not more than 3\%).
Note that integrals $I_4,\,I_6,\,I_7$ are evaluated 
analytically, and all the rest ones are calculated 
numerically. Let us recall, that only two parameters 
are in our disposal -- the average strong 
coupling constant $\bar \alpha_s$ and dimensional 
constant of $\pi$-decay $f_\pi$. We use of course 
the Goldberger-Treiman relation~(\ref{fpi}), 
which is obtained in the framework of the present 
approach. No other outside information is used. 
In particular while calculating $\sigma$-meson width 
we substitute pion mass which was obtained here in the 
corresponding approximation and was quoted in the Table. 
We see from Table 1, that the approach gives reasonable 
correspondence to the existing facts and data. We 
present information from experiments and from 
phenomenological considerations in the last right-hand 
column of the table.

Value of current mass $m_0$ is essentially larger than 
customary values. This parameter is defined by value 
$u_0$~(\ref{result1}), which is sharply dependent on 
corrections of the next orders of  $1/N$ expansion and 
on other details of the next approximations. For example, 
by a slight change in coefficient afore the second term 
in brackets in expression~(\ref{A0}) one may obtain for 
$m_0$ value being two-three times smaller than that 
presented in Table 1. At this all other parameters 
practically do not change. Taking into account this 
fact we do not consider the deviation of $m_0$ as 
critical for the approach. In comparison to the 
traditional value the modulus of quark condensate is 
essentially smaller. While its calculation we 
have seen that in the leading approximation this 
parameter is zero and the values quoted correspond to 
loop corrections. It is possible, that there are other 
contributions to the quark condensate, which give 
essential changes. For example contribution of $s$-quark 
loop to the first term in~(\ref{bqq}) evidently 
enlarge it. This problem deserves a special study. 
However just the presented values give rise to 
an interesting observation. It comes out, that 
Gell-Mann -- Renner relation~\cite{GM} agrees 
sufficiently well with the parameters of Table 1. 
Indeed In the lowest line of the table for pion mass 
we write down its values calculated by the relation
$$
m_\pi^2\,=\,-\,\frac{2\,m_0}{f_\pi^2}\,<\bar q\,q>\,.
$$
Values of $m_\pi$ for $m_0$ and $<\bar q\,q>$ from 
the table are quite satisfactory in the range of a 
reasonable accuracy. Note also that values of the quark 
condensate being lower than the traditional one are 
discussed in the literature (see e.g.~\cite{sim2}). 
Lattice 
measurements~\cite{lat} give rather wide interval of 
renormalization invariant value of the condensate 
$-<\bar q q>\, = \,((206\,\pm\,44\,\pm\,8\,\pm\,5)
\,$MeV$)^3$, in which also the values presented in Table 
1 also enter. Parameters of the $\sigma$-meson especially
its width noticeably depend on a choice of value 
$\bar \alpha_s$. Unfortunately the spread of 
data~\cite{pdg} covers all presented results. Provided 
one choose results of one work then it is possible to 
do some conclusion on preferable value of 
$\bar \alpha_s$. For example in paper~\cite{lewt} 
the processing of the set of data leads to result 
$m_\sigma\,=\,(470\, \pm\, 30)
\,$MeV,$\;\Gamma_\sigma\,=\,(590\, \pm\, 40)\,$MeV. Such 
values agree well with our results for 
$\bar \alpha_s \simeq 0.5$. At that the pion 
mass and the constituent quark mass are quite 
satisfactory.

So we may state a quantitative agreement for 
constituent quark mass and for parameters of mesons 
$\pi$ and $\sigma$, whereas for current mass and for the quark condensate we may declare at least a qualitative 
agreement. One hardly could expect more from the first 
non-perturbative approximation with only two 
parameters, one of which $f_\pi$ is strictly fixed. 
Therefore one have to admit the description of the 
low-energy data by the approach being satisfactory.

To conclude let us emphasize that the aim of the work 
is achieved. We have begun with the demonstration of 
the non-trivial solution of the compensation equation. 
The existence of scalar and pseudo-scalar excitations 
(mesons) in the same approximation is a consequence of 
its existence. The account of QCD interaction leads to 
the shift of their masses squared to the negative 
region, i.e. to the appearance of tachyons, which are 
necessary for scalar condensate to arise. As a result 
we obtain the standard scheme of chiral symmetry breaking 
with massive scalar and massless pion. Subsequent 
approximations of the approach are related to values 
of the quark condensate and the pion mass. 

We have shown that the application of the method of 
work~\cite{Arb04}, which is based on Bogolubov 
quasi-averages approach, to the low-energy region of 
hadron physics leads to quite reasonable results. From 
this we would make two essential conclusions.

Firstly, a subsequent development of the present approach 
to the hadron physics quite deserves attention. In 
particular it is advisable to apply the approach to 
calculation of parameters of vector mesons 
$\rho,\,\omega,
\,A_1$, to consider hadrons containing $s$-quark, to 
take into account the $\pi-A_1$-mixing, to introduce 
diquarks etc.. In view of methods it would be 
desirable to improve a procedure of taking into account 
of particle masses. These problems comprise subjects 
for a forthcoming studies. 

Secondly, the positive result of applicability test with 
Nambu -- Jona-Lazinio model, being taken as an example, 
allows to hope for successful application of the approach 
to other problems. In particular we mean the problem 
of a dynamical breaking of the electroweak symmetry. 
A qualitative discussion of possible variants in this 
region is presented e.g. in works~\cite{Arb1},
~\cite{SUSY}.

The author is grateful to M.K. Volkov and S.B. Gerasimov 
for valuable discussions and to I.V. Zaitsev for  
some numerical calculations.

The work is partially supported by grant "Universities 
of Russia"  UR.02.02.503.

\appendix
\section{Appendix}
We present below a set of formulas for  angular 
integrals in four-dimensional Euclidean space, 
which are used in the work.
\beq
& & x\,=\,p^2\,,\qquad y\,=\,q^2\,;\nn\\
& &\int\frac{d^4 q\,F(q^2)}{(p-q)^2}\,=\,\pi^2 
\int_0^\infty  
y dy F(y)\biggl(\theta(x-y)\frac{1}{x}+\theta(y-x)
\frac{1}{y}\biggr)\,;\nn\\
& &\int\frac{d^4 q\,F(q^2)\,(pq)}{(p-q)^2}\,=\,
\frac{\pi^2}{2} \int_0^\infty y dy F(y)\biggl(\theta(x-y)
\frac{y}{x}+\theta(y-x)\frac{x}{y}\biggr)\,;\nn\\
& &\int\frac{d^4 q\,F(q^2)(pq)^2}{(p-q)^2}\,=\,\frac
{\pi^2}{4} 
\int_0^\infty y dy F(y)\biggl(\theta(x-y)
\Bigl(\frac{y^2}{x}+y\Bigr)+\theta(y-x)\Bigl(x+\frac{x^2}
{y}\Bigr)\biggr)\,;\nn\\
& &\int\frac{d^4 q\,F(q^2) (pq)^3}{(p-q)^2}\,=\,
\frac{\pi^2}{8} \int_0^\infty 
y dy F(y)\biggl(\theta(x-y)\Bigl(\frac{y^3}{x}+2 y^2\Bigr)
+\theta(y-x)\Bigl(2 x^2+\frac{x^3}{y}\Bigr)
\biggr)\,;\nn\\
& &\int\frac{d^4 q\,F(q^2)(p,p-q)}{((p-q)^2)^2}\,=\,
\pi^2 \int_0^\infty 
y dy F(y)\theta(x-y)\frac{1}{x}\,;\nn\\
& &\int\frac{d^4 q\,F(q^2) (q,q-p)}{((p-q)^2)^2}\,=\,
\pi^2 \int_0^\infty 
y dy F(y)\theta(y-x)\frac{1}{y}
\,;\nn\\
& &\int\frac{d^4 q\,F(q^2)(p,p-q)(pq)}{((p-q)^2)^2}\,=\,
\frac{\pi^2}{4} \int_0^\infty 
y dy F(y)\biggl(\theta(x-y)\frac{3 y}{x}-\theta(y-x)
\frac{x}{y}\biggr)\,;\nn\\ 
& &\int d^4q F(q^2) \log\,(q-p)^2\;=\;\pi^2 
\int_0^\infty y dy F(y)\biggl(\theta(x-y)\Bigl
(\frac{y}{2 x}+\log\,x\Bigr)+\nn\\
& &+\,\theta(y-x)\Bigl(\log\,y+\frac{x}{2 y}\Bigr)
\biggr)\,;\nn\\ 
& &\int d^4q F(q^2) (pq) \log\,(q-p)^2\;=\;\frac{\pi^2}{6} 
\int_0^\infty y dy F(y)\biggl(\theta(x-y)\Bigl(
\frac{y^2}{x}-3 y\Bigr)+\nn\\
& &+\,\theta(y-x)\Bigl(- 3 x + \frac{x^2}{y}\Bigr)
\biggr)\,;\label{intlog}\\ 
& &\int d^4q F(q^2) (pq)^2\log\,(q-p)^2\;=\;\frac{\pi^2}
{4} \int_0^\infty y dy F(y)\biggl(\theta(x-y)\Bigl(
\frac{y^3}{4 x}
+\,x y \log\,x\Bigr)+\nn\\
& &+ \theta(y-x)\Bigl(x y \log\,y +\frac{x^3}{4 y}\Bigr)
\biggr)\,;\nn\\
& &\int d^4q F(q^2) (pq)^3 \log\,(q-p)^2\;=\;\frac{\pi^2}
{8} 
\int_0^\infty y dy F(y)\biggl(\theta(x-y)\Bigl(
\frac{y^4}{5 x}+\frac{y^3}{3}-2 y^2 x\Bigr)+\nn\\
& &+\,\theta(y-x)\Bigl(-\, 2 x^2 y + \frac{x^3}{3}+ 
\frac{x^4}{5 y}\Bigr)
\biggr)\,.\nn  
\eeq
Integrals containing $\log\,(q-p)^2$ are 
evaluated using formulas from book~\cite{Mar}.

\newpage
\begin{center}
Table 1.\\
Parameters of the low-energy hadron physics 
for three values\\
of effective coupling 
constant $\bar\alpha_s$.
\end{center}

\bigskip

\begin{center}
\begin{tabular}{|c|c|c|c|c|}
\hline

$\bar\alpha_s$ & 0.46 &  0.61 & 0.73 & 
exp/phen \\
\hline
$f_\pi\,$MeV & 93 & 93 & 93 & input\\
\hline
 $g$& 3.20 &  3.44 & 3.59 & -- 
\\
\hline
$m_0\,$MeV & 19.8 &  17.9 & 16.9 & 5 -- 10 \\
\hline
$m\,$MeV & 298.5 & 321.3 & 335.5 & 
300 -- 350 \\
\hline
 $G_1^{-1/2}\,$MeV & 240.5 &  
217.6 & 205.4  &-- \\
\hline
$-<\bar q q>^{1/3}\,$MeV 
& 0 & 0 & 0
 &   \\
 & 167.0 &  155.4 & 148.7
 & 220 -- 240  \\
\hline
$m_\pi\,$MeV
& 0 & 0 & 0
 &   \\
 & 146.0 & 124.6 & 113.3
 & 136 -- 140 \\
\hline
 $g_{\sigma\,\pi\,\pi}/m$ & 8.04  &  8.27  &  8.40 & -- \\
\hline
 $m_\sigma\,$MeV & 440.1 &  469.1 & 486.3 & 
400 -- 1200 \\
\hline
 $\Gamma_\sigma\,$MeV & 584.4 &  761.1 & 
862.5 & 600 -- 1000 \\
\hline
 $u$ & 0.001 &  0.002 & 0.003 & \\
\hline
\end{tabular}
\end{center}
\newpage
\begin{center}
{\bf Figure captions}
\end{center}
\bigskip
\bigskip
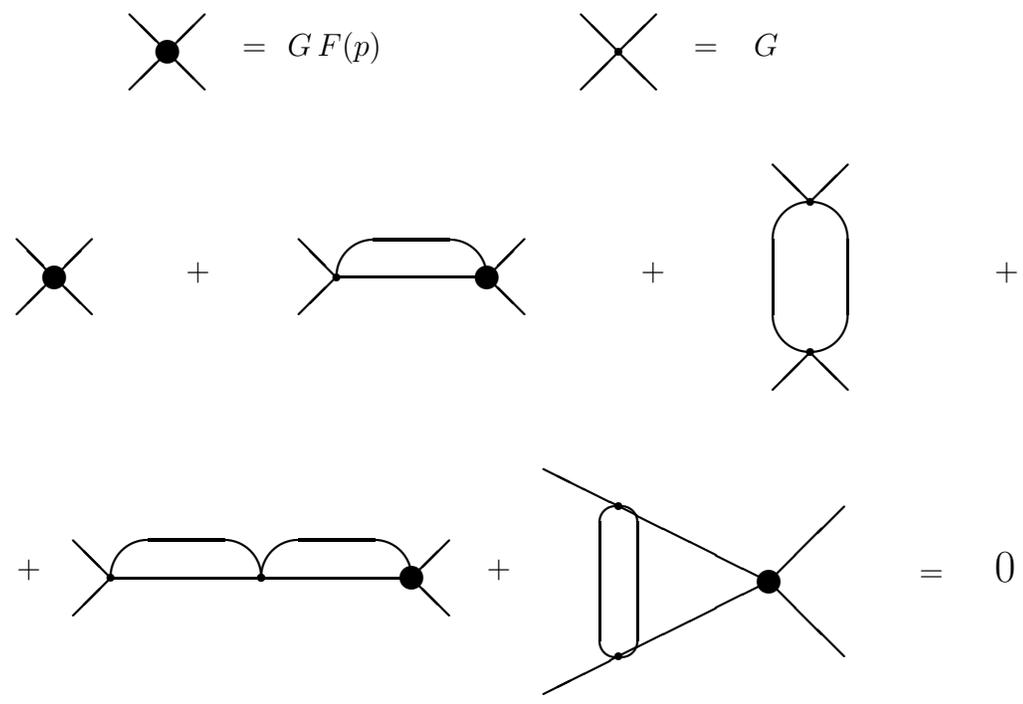
Fig. 1. Diagram representation of the compensation 
equation. Black spot corresponds to four-fermion 
vertex with a form-factor. \\
\\
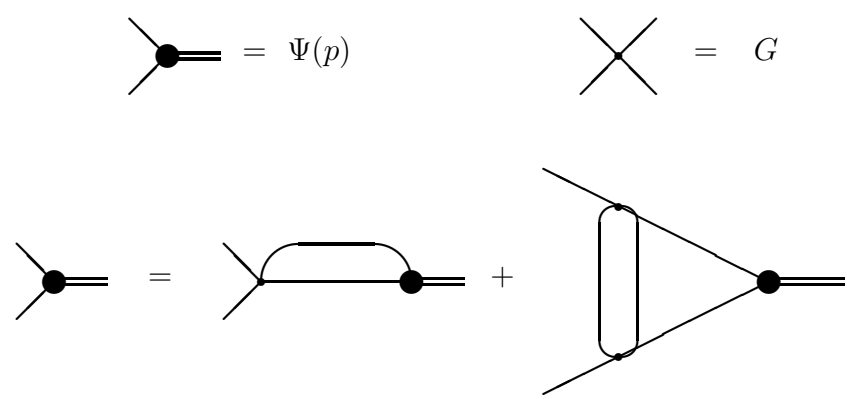
Fig. 2. 
Diagram representation of Bethe-Salpeter equation 
for scalar bound state, the later corresponding to a 
double line.\\
\\
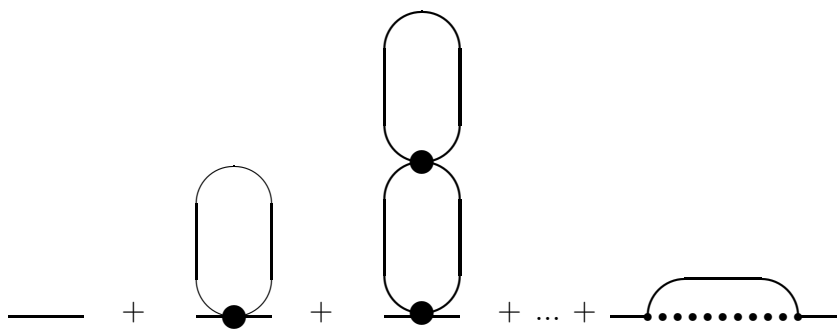
Fig. 3. Loop correction to the non-perturbative part 
of the quark condensate. The full line corresponds to 
the quark propagator with non-perturbative mass 
operator $(m-m_0)\Psi(p^2)$. The dotted line 
represents a gluon.

\newpage
\begin{picture}(160,85)
{\thicklines
\put(20,80.5){\line(-1,1){5}}
\put(20,80.5){\line(1,1){5}}
\put(20,80.5){\circle*{3}}
\put(20,80.5){\line(-1,-1){5}}
\put(20,80.5){\line(1,-1){5}}
\put(30,80){=}
\put(36,80){$G\,F(p)$}
{\thicklines
\put(80,80.5){\line(-1,1){5}}
\put(80,80.5){\line(1,1){5}}
\put(80,80.5){\circle*{1}}
\put(80,80.5){\line(-1,-1){5}}
\put(80,80.5){\line(1,-1){5}}}
\put(90,80){=}
\put(98,80){$G$}

{\thicklines
\put(5,50.5){\line(-1,1){5}}
\put(5,50.5){\line(1,1){5}}
\put(5,50.5){\circle*{3}}}
\put(5,50.5){\line(-1,-1){5}}
\put(5,50.5){\line(1,-1){5}}

\put(22.5,50){+}
{\thicklines
\put(42.5,50.5){\line(-1,1){5}} \put(52.5,50.5)
{\oval(20,10)[t]}
\put(62.5,50.5){\line(1,1){5}}\put(42.5,50.5)
{\circle*{1}}
\put(62.5,50.5){\circle*{3}}}
\put(42.5,50.5){\line(-1,-1){5}} \put(62.5,50.5)
{\line(1,-1){5}}
\put(42.5,50.5){\line(1,0){20}}
\put(83,50){+}

{\thicklines
\put(105.5,60.5){\line(-1,1){5}}
\put(105.5,60.5){\line(1,1){5}}
\put(105.5,50.5){\oval(10,20)}
\put(105.5,60.5){\circle*{1}}
\put(105.5,40.5){\circle*{1}}}
\put(105.5,40.5){\line(-1,-1){5}}
\put(105.5,40.5){\line(1,-1){5}}
\put(130,50){+}
\put(0,10.5){+}
{\thicklines
\put(12.5,10.5){\line(-1,1){5}} \put(22.5,10.5)
{\oval(20,10)[t]}
\put(12.5,10.5)
{\circle*{1}}
\put(32.5,10.5){\circle*{1}}}
\put(12.5,10.5){\line(-1,-1){5}} 
\put(12.5,10.5){\line(1,0){20}}
{\thicklines
 \put(42.5,10.5)
{\oval(20,10)[t]}
\put(52.5,10.5){\line(1,1){5}}\put(32.5,10.5)
{\circle*{1}}
\put(52.5,10.5){\circle*{3}}}
 \put(52.5,10.5)
{\line(1,-1){5}}
\put(32.5,10.5){\line(1,0){20}}
\put(62.5,10.5){+}
{\thicklines
\put(100,10){\line(-2,1){30}}
\put(100,10){\line(-2,-1){30}}
\put(80,10){\oval(5,20)}
\put(80,20){\circle*{1}}
\put(80,0){\circle*{1}}
\put(100,10){\line(1,1){10}}
\put(100,10){\line(1,-1){10}}
\put(100,10){\circle*{3}}}}
\put(120,10){=}
\put(130,10){{\Large 0}}
\end{picture}

\bigskip
\bigskip
\bigskip
\bigskip
\bigskip
\bigskip
\bigskip
\bigskip
\bigskip
\bigskip
\bigskip
\bigskip
\bigskip
\bigskip
\bigskip
\bigskip
\bigskip
\bigskip

\begin{center}
Fig. 1. 
\end{center}

\newpage
\begin{picture}(160,85)
{\thicklines
\put(20,50.5){\line(-1,1){5}}
\put(20,50.5){\circle*{3}}
\put(20,50.5){\line(-1,-1){5}}
\put(20,50.9){\line(1,0){7}}
\put(20,50.1){\line(1,0){7}}
\put(30,50){=}
\put(36,50){$\Psi(p)$}
{\thicklines
\put(80,50.5){\line(-1,1){5}}
\put(80,50.5){\line(1,1){5}}
\put(80,50.5){\circle*{1}}
\put(80,50.5){\line(-1,-1){5}}
\put(80,50.5){\line(1,-1){5}}}
\put(90,50){=}
\put(98,50){$G$}

{\thicklines
\put(5,20.5){\line(-1,1){5}}
\put(5,20.5){\circle*{3}}}
\put(5,20.5){\line(-1,-1){5}}
\put(5,20.9){\line(1,0){7}}
\put(5,20.1){\line(1,0){7}}
\put(17.5,20){=}
{\thicklines
\put(32.5,20.5){\line(-1,1){5}} 
\put(42.5,20.5){\oval(20,10)[t]}
\put(52.5,20.9){\line(1,0){7}}
\put(32.5,20.5)
{\circle*{1}}
\put(52.5,20.5){\circle*{3}}}
\put(32.5,20.5){\line(-1,-1){5}} 
\put(52.5,20.1){\line(1,0){7}}
\put(32.5,20.5){\line(1,0){20}}
\put(63,20){+}

{\thicklines
\put(100,20.5){\line(-2,1){30}}
\put(100,20.5){\line(-2,-1){30}}
\put(80,20.5){\oval(5,20)}
\put(80,30.5){\circle*{1}}
\put(80,10.5){\circle*{1}}
\put(100,20.9){\line(1,0){10}}
\put(100,20.1){\line(1,0){10}}
\put(100,20.5){\circle*{3}}}}

\end{picture}
\bigskip
\bigskip
\bigskip
\bigskip
\bigskip
\bigskip
\bigskip
\bigskip
\bigskip
\bigskip
\bigskip
\bigskip
\bigskip
\bigskip
\bigskip
\bigskip
\bigskip

\begin{center}
Fig. 2. 
\end{center}
\newpage
\begin{picture}(160,55)

{\thicklines
\put(5,0.5){\line(1,0){10}}}
\put(20,0){+}
{\thicklines
\put(35,0.5){\circle*{3}}}
\put(30,0.5){\line(1,0){10}}
\put(35,10.5){\oval(10,20)}}
\put(45,0){+}
{\thicklines
\put(55,0.5){\line(1,0){10}
\put(-5,0.5){\circle*{3}}
\put(-5,20.5){\circle*{3}}
\put(-5,10.5){\oval(10,20)}
\put(-5,30.5){\oval(10,20)}}
\put(70,0){+}
\put(75,0){...}
\put(80,0){+}
{\thicklines
\put(85,0.5){\line(1,0){5}}
\put(110,0.5){\line(1,0){5}}
\multiput(90,0.5)(2,0){11}{\circle*{1}} 
\put(100,0.5){\oval(20,10)[t]}}
\end{picture}

\bigskip
\bigskip
\bigskip
\bigskip
\bigskip
\bigskip
\bigskip
\bigskip
\bigskip
\bigskip
\bigskip
\bigskip
\bigskip
\bigskip
\bigskip
\bigskip
\bigskip
\bigskip
\bigskip
\bigskip
\begin{center}
Fig. 3. 
\end{center}
\newpage

\end{document}